\documentclass[sigconf]{acmart}

\usepackage{float}
\usepackage{subfigure}
\AtBeginDocument{%
  \providecommand\BibTeX{{%
    \normalfont B\kern-0.5em{\scshape i\kern-0.25em b}\kern-0.8em\TeX}}}
\usepackage{tabularx}
\usepackage{bbding}
\usepackage{graphicx}
\usepackage{geometry}
\usepackage{subfigure}
\usepackage{amsmath}
\usepackage{makecell}
\usepackage{float}
\usepackage{color,soul}
\usepackage{booktabs}
\usepackage{caption}
\usepackage{tabu}
\usepackage{hyperref}

\copyrightyear{2025}
\acmYear{2025}
\setcopyright{rightsretained}
\acmConference[CUI '25]{Proceedings of the 7th ACM Conference on
Conversational User Interfaces}{July 8--10, 2025}{Waterloo, ON, Canada}
\acmBooktitle{Proceedings of the 7th ACM Conference on Conversational User
Interfaces (CUI '25), July 8--10, 2025, Waterloo, ON, Canada}
\acmDOI{10.1145/3719160.3735658}
\acmISBN{979-8-4007-1527-3/2025/07}

\begin{document}

\title[Eternagram]{Eternagram: Inspiring Climate Action Through LLM-based Conversational Exploration of a Post-Devastation Climate Future}


\author{Suifang Zhou}
\orcid{0000-0001-5103-580X}
\authornote{Correspondence should be addressed to LC@raylc.org}
\affiliation{%
  \institution{Studio for Narrative Spaces\\City University of Hong Kong}
  \country{Hong Kong}}
\email{sfzhou3-c@my.cityu.edu.hk}

\author{RAY LC}
\orcid{0000-0001-7310-8790}

\affiliation{
\institution{Studio for Narrative Spaces\\City University of Hong Kong}
\country{Hong Kong}}
\email{LC@raylc.org}

\renewcommand{\shortauthors}{LC}

\begin{abstract}
Climate action is difficult to persuade because we tend to perceive climate change as remote and disconnected from daily life. Instead of traditional informational engagements, game-based interventions can create narratives that immerse the visitor in situations where their actions have tangible consequences. To make these narratives engaging, we used a speculative scenario of an alien stumbling upon social media to obliquely address climate change through a text-based adventure game installation. Mimicking visitors’ natural dialogue
in social media apps, we designed an LLM-based chatbot with knowledge of post-climate devastated world that mirrors our own planet Earth. In discovering the world’s downfall through interactive chatting and posted images, players begin to realize that their own actions can make a difference on impacts of climate change in this distant world, fostering pro-environmental attitudes. Previously published at CHI, this game installation demonstrates the potential of LLM-based creative narratives in exploring speculative worlds driving social change.
\end{abstract}



\begin{teaserfigure}
  \includegraphics[width=\textwidth]{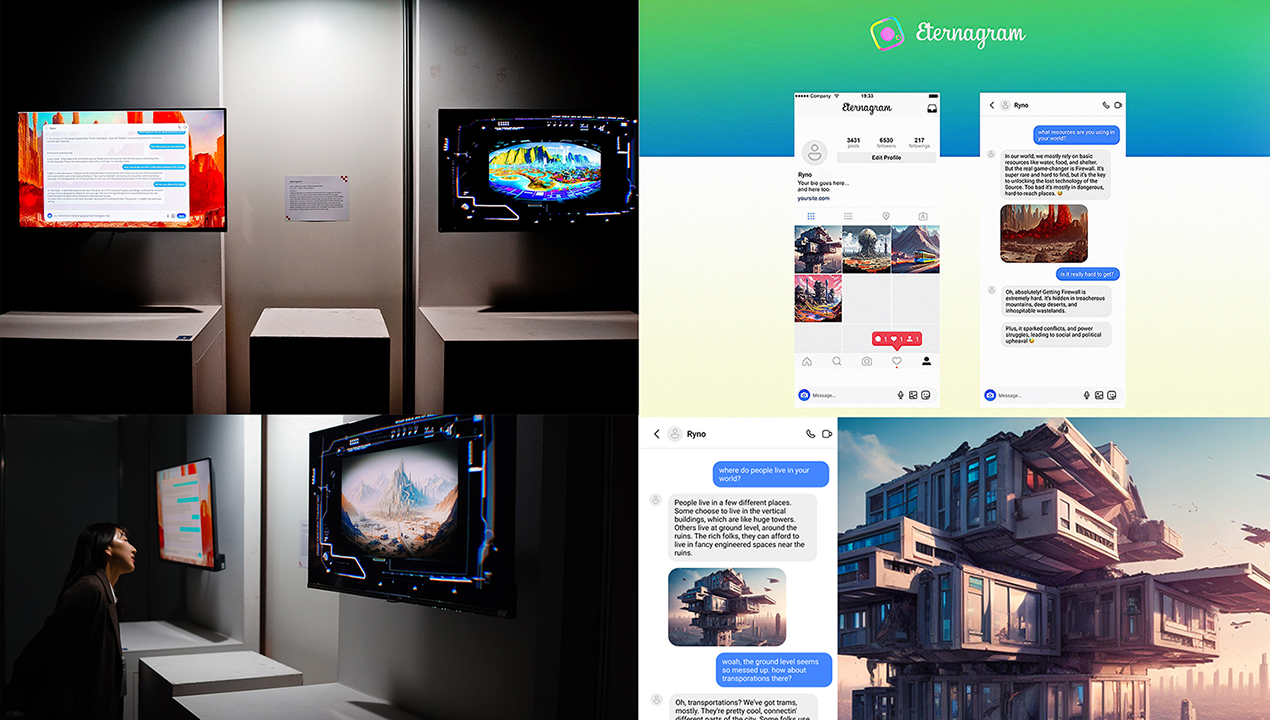}
  \caption{Eternagram as physical installation and social media game. \textbf{(Left)} Eternagram shown as two videos, one showing the dialogues with the system on social media, and one as image animation depictions, using two screens and pedestals. \textbf{(Right)} Eternagram as interactive app that can be accessed online.}
  \Description{Eternagram as physical installation and social media game. (Left) Eternagram shown as two videos, one showing the dialogues with the system on social media, and one as image animation depictions, using two screens and pedestals. (Right) Eternagram as interactive app that can be accessed online.}
  \label{fig:teaser}
\end{teaserfigure}

\maketitle

\section{Introduction}\label{sec:Intro}
We have been considering climate change as a physical phenomenon that demands policy and resource-level physical solutions ~\cite{lc_designing_2022, lc_speculative_2024}, but the true agent for change may lie inside ourselves on the human level ~\cite{lc_designing_2021}. Promoting pro-climate behavioral change faces challenges because such issues are often perceived as distant in time, irrelevant to immediate needs, and beyond our control, seemingly far removed from daily life ~\cite{lc_designing_2022, song_climate_2021,lc_chikyuchi_2022}. Therefore, pro-climate creative interventions must first engage with people's intrinsic motivations on a mental and community level, using creative solutions to align people with the goals of climate action.

One such creative engagement that can align people with actionable change for pro-climate behaviors are interactive games ~\cite{zhang_can_2025,li_generative_2024}. Exhibitions involving games can provide visitors avenues to discover the world around them and align potential actions with their intrinsic motives ~\cite{zhou_eternagram_2024,agcal_bricolage_2025}. To engage public attention and affect both those who do, and do not align with climate action, we immerse visitors in a conversational text adventure game installation ~\cite{zhou_eternagram_2024-1}, where visitors can explore and discover a post-climate devastation world through a social-media interface that feels close to their daily lives. They are enabled to engage in conversations with an LLM-based mysterious stranger who has lost her memory and accidentally joined our world’s social network.

As players try to help recover her memory, they slowly begin to unravel the stories of a distant world by conversing with this stranger through the application entitled Eternagram. They learn about the mysteries of this world by viewing the stranger's shared images and videos, which are generated using world-building statements as prompts for a custom Stable Diffusion model ~\cite{liu_dreamscaping_2024,liu_falling_2024,li_affecting_2024}. Players soon see the eerie similarities between the history of this distant planet and our own Earth, discovering how the demise of the stranger's world occurred and drawing lessons for what we can do here on Earth. The social-media and free-conversation format of the installation connects the audience’s everyday lives to a speculative future in the form of a distant world~\cite{wu_present_2024,lc_imitations_2021,lc_imitations_2022,fu_being_2024,he_i_2025,lc_speculative_2024}, using ChatGPT to create a believable character that players can relate to, learn from, and ultimately, align their own behaviors with ~\cite{han_when_2024,yang_ai_2022,zeng_ronaldos_2025}. Such LLM-based interactions provide natural language interfaces to immerse the audience in discussion, argumentation, and reflection ~\cite{zeng_ronaldos_2025} designed to creatively engage visitors with the narrative. The game-art story then provides awareness and reflection for a speculative world that becomes aligned to real world actionable behaviors that players learn to internalize. Indeed, the climate awareness metrics of players post-game was previously found to be correlated to in-game perceptions and attitudes~\cite{zhou_eternagram_2024}, thus providing a theoretical basis for the potential effectiveness of the installation work to foster climate thinking and intention to align behaviors to climate action.

\section{Art Work Description}\label{sec:Description}
Eternagram is presented in three modular forms including: 

\textbf{(1)} An online interactive interface designed to emulate a social network, enabling audiences to engage in text-based conversations with a chatbot online live. Visitors can play online on their own phones.

\textbf{(2)} The corresponding images and animations that illustrate the in-game climate scene generated by Stable Diffusion models shown in video form. The image video is also available on youtube.

\textbf{(3)} A text flow of the story of the interaction shown through an example interaction with the AI-based character. The visitor can read slowly through a concise video of the main points in the conversational chat dialogue highlighting the story.

Eternagram integrates diverse data streams and narrative elements to engage audiences emotionally and psychologically. This text adventure game employs prompt engineering and a custom text database to render post-climate game scenes. This approach allows audiences to engage in data-driven storytelling through textual narratives and visual content, presenting complex climate change information in an accessible format. Stable Diffusion models with fine-tuned parameters generate images that visualize the consequences of climate change, making this abstract concept tangible and relevant to individuals' daily lives.

Previous projects on human-GenAI interactions have demonstrated that audiences find these experiences engaging  ~\cite{lc_presentation_2022,lc_contradiction_2023,cao_dreamvr_2023,lc_active_2023}, particularly in bridging the affective gap between personal experiences and abstract, global challenges such as climate change ~\cite{lc_together_2023, lc_time_2024, lc_human_2023, lc_imitations_2022}. Eternagram leverages the natural language ability of GenAI-supported gameplay~\cite{ling_sketchar_2024,sun_bringing_2022} to invite participants to immerse themselves in an interactive narrative based on a social-media-like interface, feeling as though they are involved in real life. Through playful interactions by text with the AI-based agent, visitors can explore an alternative, futuristic, yet reality-related climate scenario while linking this future with their social media present. Thus, we attempt to foster a deeper understanding and engagement with climate change in the real world by linking real-life social media interactions with a larger-than-life speculative world by the way of a custom prompted GenAI chat agent.

\subsection{Design for GPT-powered Interactive Game Storytelling}

In conventional narrative-focused games, player exploration of the game world is often facilitated through structured dialogue systems. These typically involve predefined narrative and dialogue choices for interactions between the player and non-player characters (NPCs). A key advantage of this framework is that designers retain control over the range of permissible player responses and the corresponding feedback mechanisms. However, such dialogue systems can be analogized to dropdown menus: players are limited to selecting from options that have been preordained by the designer. This inherently constrains the spontaneity and depth of exploratory behavior within the game world.

\begin{figure*}[htbp]
 \centering
   \includegraphics[width=\textwidth]{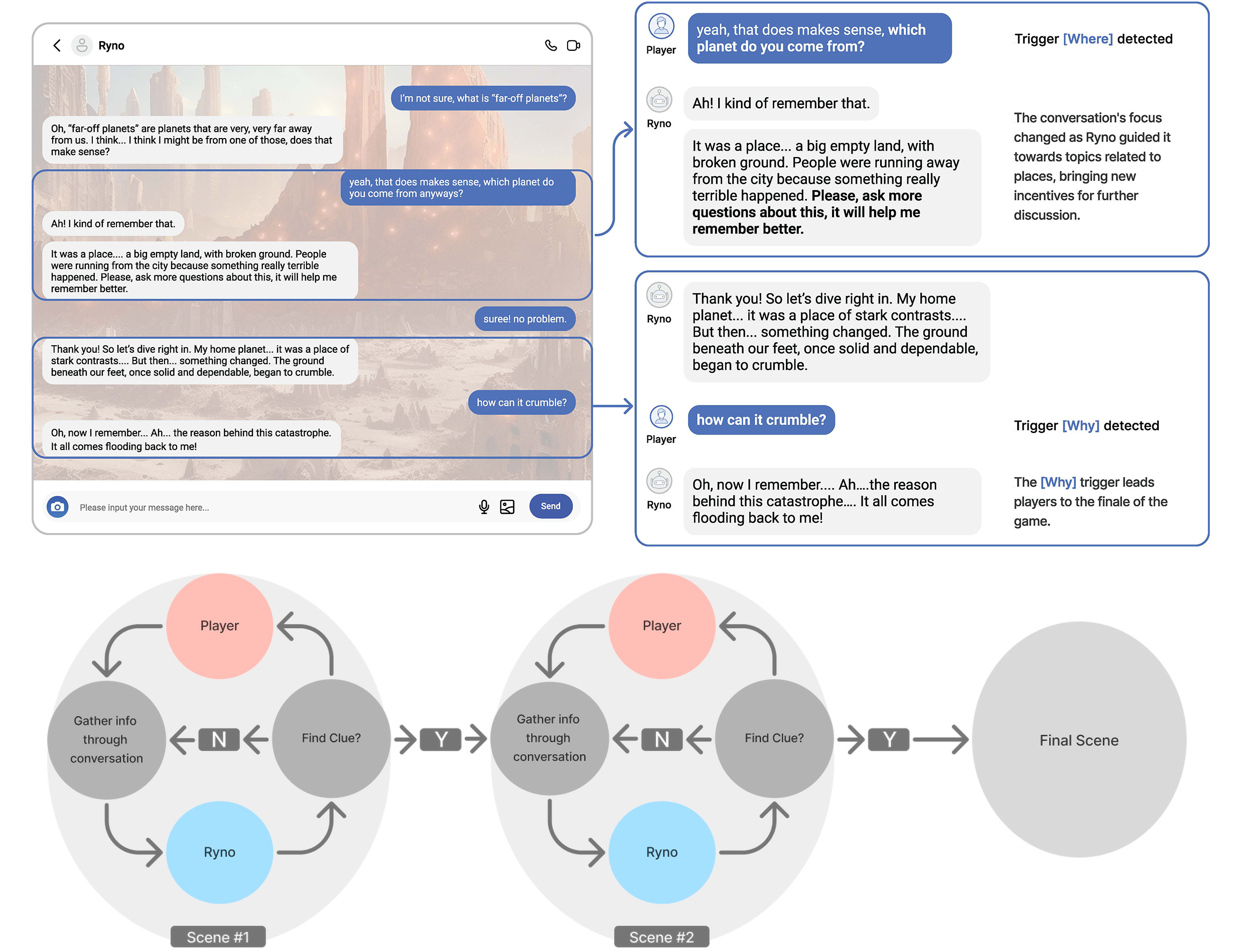}
   \caption{Example game flow for the interactive form. \textbf{(Top)} How players can trigger the GenAI agent to go from one place to another in the conversation as they explore the world through natural conversation. \textbf{(Bottom)} The way the full game flow works in the interactive online version. Players can stop at any point but have the option to explore the entire story by making it to the final scene.}
   \label{fig:flow}
   \Description{Example game flow for the interactive form. (Top) How players can trigger the GenAI agent to go from one place to another in the conversation as they explore the world through natural conversation. (Bottom) The way the full game flow works in the interactive online version. Players can stop at any point but have the option to explore the entire story by making it to the final scene.}
 \end{figure*}

In contrast, the present game introduces an experimental ludonarrative mechanic that enables free-form dialogue between the player and a GPT-powered NPC (Figure ~\ref{fig:flow}), aiming to foster a novel mode of interaction and storytelling. This approach invites players to engage with the game narrative in a more open-ended manner, emphasizing discovery and improvisation rather than predetermined choice.

The implementation of this free-form dialogue system relies on three key components: a narrative corpus representing the game world, a game motivation or goal that instills suspense and drives progression, and Ryno—a carefully prompted GPT-powered NPC whose persona is intricately tied to the game’s storyline. Player exploration of the game world unfolds primarily through conversational interaction with Ryno.

\subsubsection{Design of Game Narrative Corpus} The game’s narrative corpus comprises two primary categories of information: foundational world knowledge and trigger-related insights. The foundational knowledge is drawn from a carefully curated world-building text corpus, depicting a science fiction universe shaped by the aftermath of climate catastrophe. This corpus offers players a comprehensive understanding of the game’s setting, including its historical timeline, sociopolitical structures, technological landscape, cosmology, and the forms of life that inhabit the world. It details both the events leading up to the ecological collapse and the transformations that followed, providing rich contextual material for immersive exploration. In contrast, trigger-related insights are specific narrative cues that, once identified and articulated by the player, activate key progression mechanisms—such as prompting the NPC to disclose its origin. These narrative triggers act as milestones that structure the player’s journey through the game, marking transitions between gameplay phases and introducing new level-specific goals.

\subsubsection{Design of Game Motivation} The game is structured around a central overarching objective—assisting a mysterious online stranger in recovering his lost memories—introduced during the prologue to provide long-term narrative motivation. Supporting this main goal is a series of intermediate level-specific objectives, which are revealed through end-of-level cutscenes that precede each new gameplay phase. These level-goals function to regulate narrative pacing, emphasize the game’s layered design, and sustain player engagement across different stages. Conceptually, the game resembles an evolving text-based puzzle adventure, casting the player in the role of a "detective" who actively uncovers scattered clues through open-ended dialogue. As players progress, they gather narrative insights by triggering predefined events and gradually deepening their understanding of the game world. This accumulated knowledge ultimately equips them to resolve the central mystery and fulfill the game’s main objective.

\subsubsection{Design of Ryno} Our system employs GPT-4 to power the character of Ryno. Within the narrative, Ryno serves a central role by engaging players in continuous dialogue, evaluating their responses for narrative relevance, and determining whether sufficient information has been uncovered to permit progression. Ryno retains memory of previous conversations with the player, allowing for context-aware interactions that evolve over time. His dialogue is shaped by both the world-building corpus and the accumulated dialogue history, facilitating consistency in characterization. To align with the overarching narrative theme, Ryno’s persona is crafted as a mysterious figure who has lost his memory, thereby encouraging players to reconstruct his identity through dialogue—a process that parallels their journey of uncovering the broader game world.

\subsection{Game Interface Design}

\begin{figure*}[h]
    \centering
    \includegraphics[width=\textwidth]{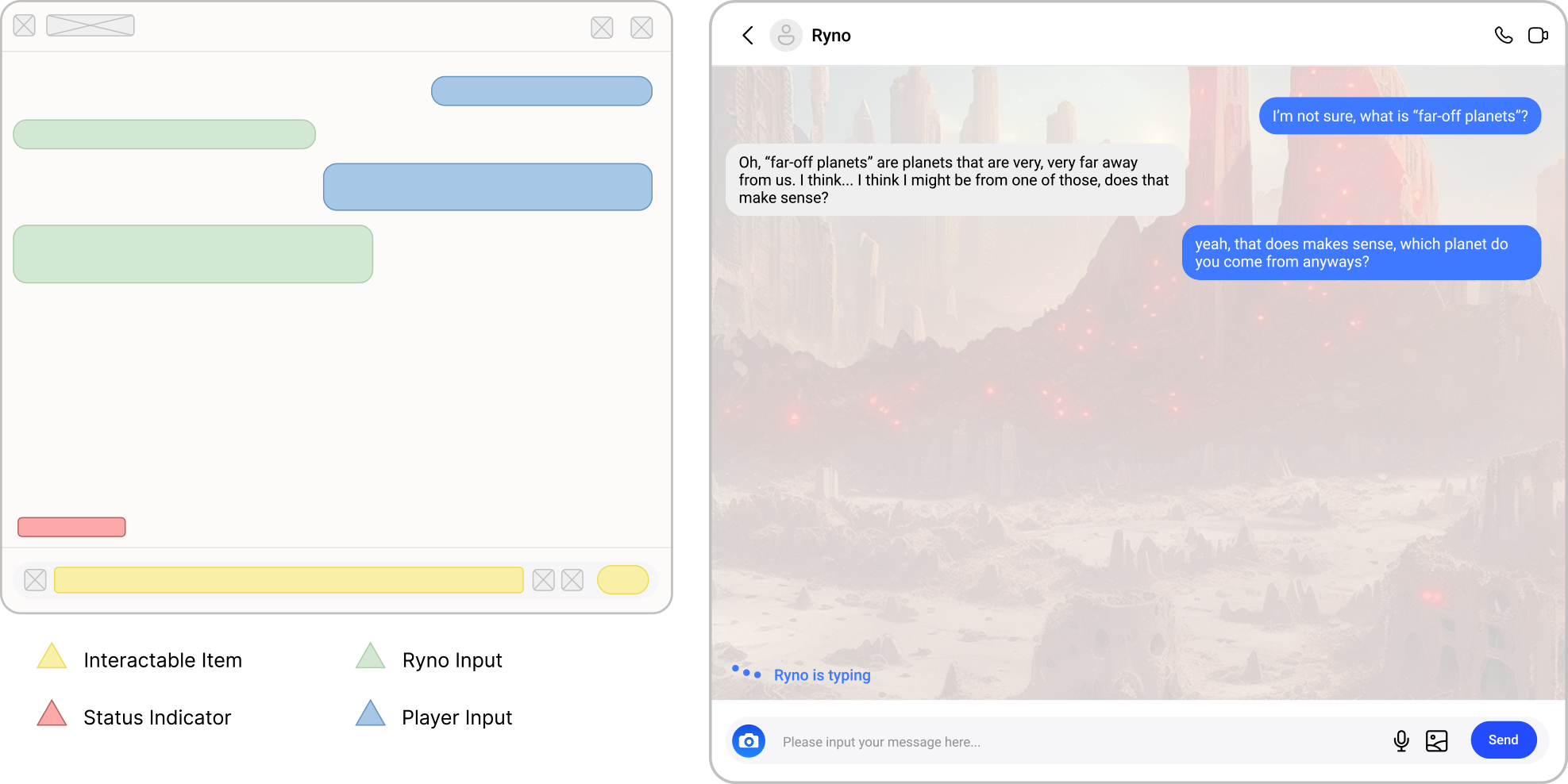}
    \caption{The in-game interface emulates the original design of Instagram's UI}
    \label{fig:interface}
\end{figure*}

The game's setting mimics a social networking platform (Figure~\ref{fig:interface}). Our motivation for this design is to address the question of how to allow players to express attitudes toward climate issues in a similar way to their expression in real-world contexts. We reasoned that the social network style allows players to immerse themselves in the game and behave as they typically would, since it looks just like their own daily use. Even as players gradually uncover a profound and unexpected storyline, the interactions facilitating this revelation remain rooted in mundane, everyday experiences akin to chatting on social media.

Also reflected in the game UI is the deliberate game aesthetic that takes players from the ordinary connection to their own lives to the happenings in a strange new land. The social media interface is designed to lead players to perceive the GPT character as another regular individual conversing with them online. Meanwhile, the progressive infusion of fantastical elements within this reality-based setting, such as sudden revelations of memories and evolving story arcs, evokes a sense of the surreal seamlessly infiltrating the real world.

\subsection{Gameplay Walkthrough}

\begin{figure*}[htbp]
 \centering
   \includegraphics[width=\textwidth]{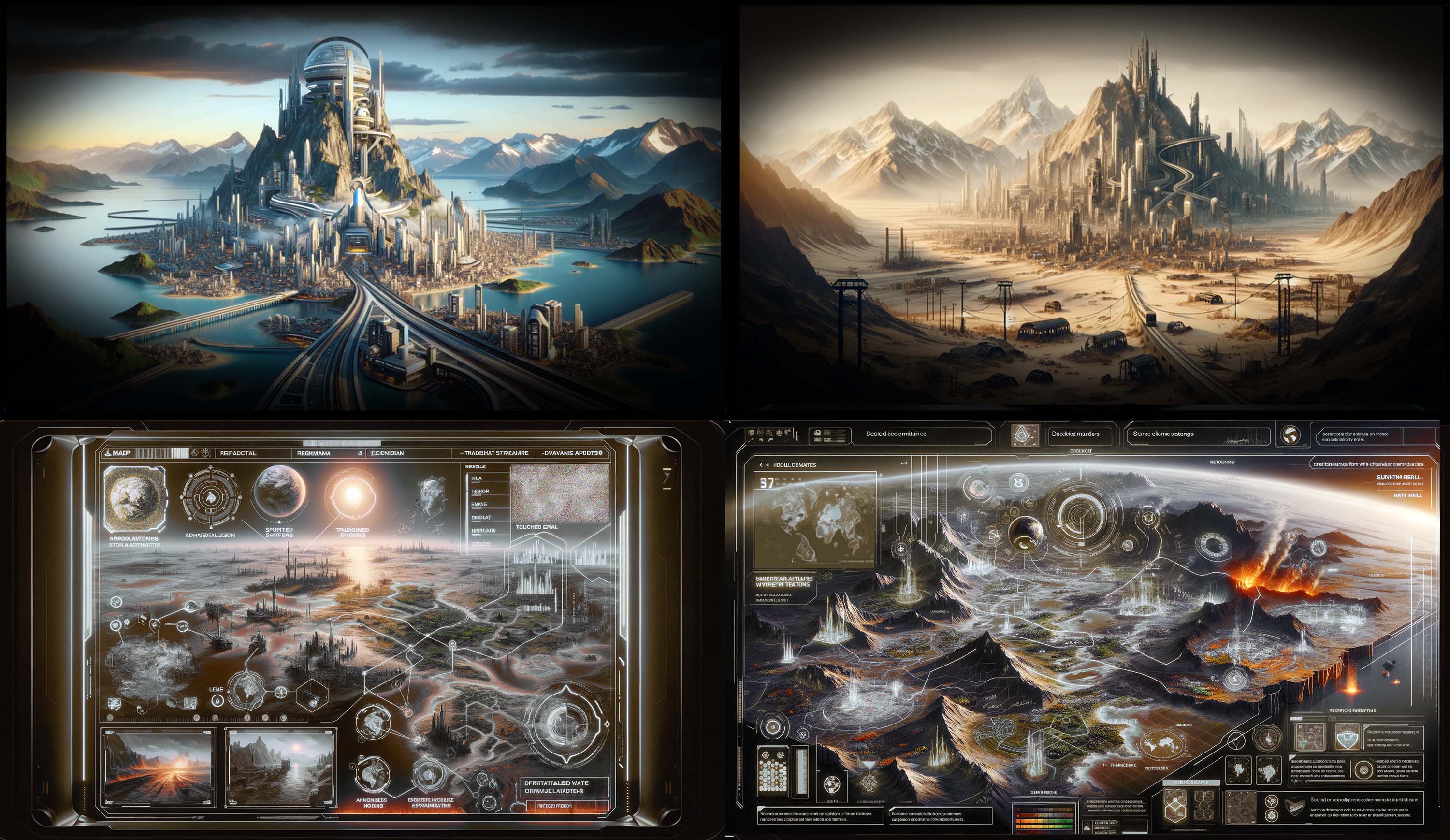}
   \caption{Narrative engendered by the conversation-based interaction shown as GenAI-created images. Through interactive dialogue with the stranger, the player uncovers \textbf{(Left)} a once-thriving, once advanced civilization that has now fallen into ruin \textbf{(Right)} as described by the LLM-based chatbot character due to a sequence of events leading to catastrophic climate change.}
   \label{fig:beforeafter}
   \Description{Caption}
 \end{figure*}

From the visitor's perspective, the gameplay unfolds as follows: 

On an otherwise ordinary day of browsing online social platforms, the player encounters a mysterious individual named Ryno suffering from amnesia, seeking their assistance. The core objective of the game revolves around the player’s interactions with this enigmatic stranger as they attempt to restore the Ryno's lost memories through text-based exchanges.

As the dialogue unfolds across multiple rounds, the player begins to sense that Ryno is far from ordinary. They appear to hail from a distant, unsettling place ravaged by severe environmental crises. As the exchanges deepen, it becomes evident that recovering the stranger’s memories is intrinsically linked to restoring the very world tied to their fragmented recollections.

With each successful revelation, the player uncovers more secrets about an entrancing extraterrestrial civilization—one drastically different from Earth, yet similar in some ways. Through these conversations, the player pieces together a post-human-like existence, where inhabitants are forced to live underground to escape extreme weather. To conserve dwindling energy resources across generations, some have resorted to prolonged hibernation in virtual reality, as digital existence has become the most efficient way to sustain life without consuming excessive energy. A once-thriving city now lies in ruins (Figure~\ref{fig:beforeafter}), its history marred by ecological devastation that reshaped both its landscapes and its people.

Ultimately, as the stranger’s true identity comes to light, the player reaches a startling realization: the entity they have been communicating with is neither human nor organic. This is no mere online encounter—it is an anomaly. The stranger is, in fact, an archive-based AI, created by an advanced civilization as an oversight bot, tasked with monitoring the socio-environmental parameters of its dying planet. Somehow, Ryno has become entangled in Earth's digital networks, reaching out in an unintentional plea for help.


\section{Proposed Installation}\label{sec:Methods}
\subsection{Physical Installation}

\begin{figure*}[htbp]
 \centering
   \includegraphics[width=\textwidth]{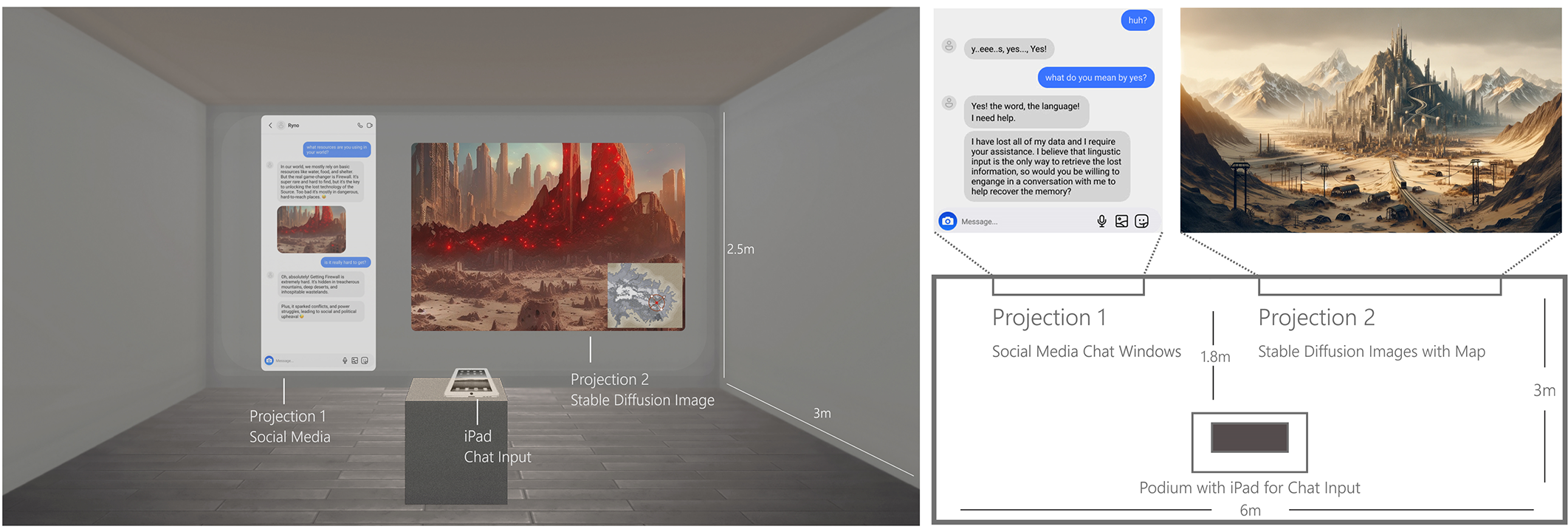}
   \caption{Floor plan and layout for a potential physical installation, including 3D view with setup needed \textbf{(Left)} and detailed specs and measurements recommended \textbf{(Right)}.}
   \label{fig:physical}
   \Description{Floor plan and layout for a potential physical installation, including 3D view with setup needed (Left) and detailed specs and measurements recommended (Right).}
 \end{figure*}

The physical installation (for subsequent years) involves: (1) a text-based adventure game and (2) Stable Diffusion-generated artwork that visually represents the in-game climate scenarios. 

The installation will feature clear and unobstructed projections. These projections will display a chat window on the left and Stable Diffusion-generated images on the right. An iPad placed on a podium within the viewing space will allow audience members to input text (Figures \ref{fig:physical}, \ref{fig:teaser}). Primary interaction occurs through audience members typing and chatting with a chatbot on an iPad. Participants observe the conversation on a left-side projection and corresponding images on the right.

To optimize an engaging visit for audiences, the videos are designed to show the most interesting elements of the conversation that leads to a full grasp of the basic premise of the story in the chat window, as well as the visuals from GenAI that correspond to the story being told. Some visitors may only look at these two videos as they scan through the exhibition space. However for those with more initiative for exploring the conversational user interface itself, we present the ipad for them to dwell into the specific story segments by interactive chat with the AI main character Ryno. Using the ipad connected to the LLM-based interaction on the website, the visitor can explore the specific character and the world that is told through its conversation. Thus the installation offers both casual and motivated visitors ways to explore the world of Eternagram, either through passive viewing or by conversational interaction. 

\subsection{Technical and Safety Information}

For online participation, visitors needs to have an internet connection with ability to join the web conversational chat. They will have the option to view the videos of the game play and generative animations at the installation site, or access the Eternagram website directly to play the game there. While the character in the game does not use obscene or inappropriate language that we know of, it is based on a prompted GenAI model, and there is always possibility of unforseen language being used that may be offensive to particular audiences, for example of particular political persuasions.

For the physical installation, the following equipment is necessary:
\begin{itemize}
\item a darkened space of around 3x3 meters in surface area.
\item 2 projectors of greater than 1000 lumens each or 2 LCDs of 36 inches or greater.
\item 1 ipad or 1 iphone with access to wifi internet at the exhibition location.
\item 1 white pedestal with surface area of 32x32 cm and around 76 cm tall.
\item dim lighting around the pedestal area of slightly diffused warm-colored spotlight.
\item 3 electrical outlets for the projection or LCDs, and for continuous charging of the mobile devices hidden inside the pedestal.
\item (optional) 2 speakers of around 10 W each for the animation sound.
\end{itemize}

Visitors should be aware that GenAI model outputs can infrequently say things that may be offensive even though we did not prompt it to do so. They are enabled to interact with the ipad in typing responses to the character in the game but is not required to do so. The text in the video game play on the projection will be made accessible to most viewers by determining the right distance and resolution of the display. The animation should not cause any aversive reactions, but there will be occasional blinking lights during the transitions between animations. Pleas also note that low carbon prompts were used as much as possible during the prototyping for the creation of this work. Moreover, a publicly available Stable Diffusion 2.0 engine was used for the creation of the animation images to maximize sustainability.

\section{Links and Videos}\label{sec:Portfolio}
\begin{itemize}
\item Online game link: \url{https://chat-climate-game-ten.vercel.app/}
\item Exhibition video: \url{https://youtu.be/GojK6OPLLK8?t=52}
\item Project website: \url{https://recfro.github.io/eternagram/}
\end{itemize}

\section{Contribution}\label{sec:Summary}
Eternagram addresses how we can adapt to climate futures and take actions for individual sustainable and responsible social purpose by the form of a speculative world where climate destruction occurred told through conversational interactions with a GenAI-based character. Visitors are enabled to explore creatively through individual dialogue with this tailored AI character, who narrates the story of its planet and its history, but this story eventually hits close to home, resembling that of Earth. These narratives are told both in the form of video and images, and also through a weblink where visitors can experience the story for themselves through conversation with the ChatGPT-based agent. Eternagram is based on previously published foundations of a CHI publication ~\cite{zhou_eternagram_2024} and VINCI exhibition ~\cite{zhou_eternagram_2024-1}, aiming to diversify its audience to the more general public for expressing how creative exploration and design may align with, and inspire social good. In apply GenAI to empower design\cite{liu_salt_2025,zhou_retrochat_2025}, it illustrates how conversational user interfaces can be used for speculative world building and game play in the context of interactive exhibition engagements.

\bibliographystyle{ACM-Reference-Format}
\bibliography{references}

\end{document}